\documentclass[aps,preprint]{revtex4}
\usepackage{epsfig}

\begin{document}

\title{Phase transition in 
conservative diffusive contact processes}

\author{Carlos E. Fiore and M\'{a}rio J. de Oliveira}

\affiliation{Instituto de F\'{\i}sica\\
Universidade de S\~{a}o Paulo\\
Caixa Postal 66318\\
05315-970 S\~{a}o Paulo, S\~{a}o Paulo, Brazil}
\date{\today}

\begin{abstract}

We determine the phase diagrams of conservative diffusive
contact processes by means of numerical simulations. 
These models are versions of the ordinary diffusive
single-creation, pair-creation and triplet-creation contact processes
in which the particle number is conserved.  
The transition between the frozen and active states
was determined by studying the system in the subcritical
regime and the nature of the transition, whether continuous or
first order, was determined by looking at the
fractal dimension of the critical cluster. 
For the single-creation model the transition
remains continuous for any diffusion rate. For pair-
and triplet-creation models, however, the transition
becomes first order for high enough diffusion rate.
Our results indicate that in the limit of infinite diffusion rate the 
jump in density equals 2/3 for the pair-creation
model and 5/6 for the triplet-creation model.


\end{abstract}

\maketitle

\section{Introduction}

The usual contact process \cite{harris,liggett, marro} is a 
model for nonequilibrium process
composed by two subprocesses: a catalytic creation
and a spontaneous annihilation. In the usual contact process, 
which we call single-creation contact process, a particle
is created if the site has at least one 
neighbor site occupied.
Many generalizations \cite{dic89a,dic89b,dic02} can be made by taking
into account the basic mechanisms 
contained in the single-creation model.
In the pair-creation contact process, for instance,
at least two neighbor sites occupied are necessary to create
a new particle. In the triplet-creation model
one should have at least three sites occupied. All these variants of
the contact process exhibit a continuous phase transition
between an absorbing state and an active state that belongs
to the direct percolation (DP) universality class.

Diffusive models are defined by permitting the diffusion
of particles in addition to the catalytic creation
and spontaneous annihilation. A diffusion process is done 
by moving a particle to an empty nearest neighbor site.
The introduction of diffusion does not destroy
the transition from an active state to an absorbing state 
giving rise to a transition line that separates the two phases.
Jensen and Dickman \cite{jen1993} have shown that
for the single-creation diffusive contact process
this line is always second order for any diffusion rate and 
belongs to the DP universality class, i. e., 
the addition of the diffusion
does not change the universality class nor the nature of transition.
For the pair-creation the numerical results by
Dickman and Tom\'e \cite{dictome}
show that the transition is continuous 
for low diffusion and they argue that there is no
change in the nature of the transition for high diffusion.
For the triplet-creation contact process, 
Dickman and Tom\'e \cite{dictome} have shown that the transition 
becomes a first order for high enough diffusion.

In the present work, we study conservative versions
of the models mentioned above. 
A conservative version of a model for nonequilibrium
process was introduced by Ziff and Brozilow \cite{brziff},
who used a constant-density ensemble to study the ZGB model. 
A conservative version
of directed percolation was used by Br\"oker and Grassberger
\cite{brograss}.
The conservative contact process was introduced by Tom\'e
and de Oliveira \cite{tome2001} who have shown its equivalence in the
thermodynamic limit to the ordinary contact process
and how to calculate the 
rates from averages over the constant-density ensemble.
The equivalence between this ensemble and the constant-rate ensemble
was later proved by Hilhorst and Wijland \cite{hilhorst}.

In the conservative versions of contact processes
\cite{tome2001,sabag,mario2003}, 
an empty site becomes occupied in a way similar 
to the catalytic creation. But instead of creating a new
particle, like in the ordinary contact processes,
a randomly chosen particle of the system leaves its
place and jumps into the empty site. Thus, both the
creation and annihilation processes are replaced
with a jumping process. However, this is not an
unrestricted jumping because particles are not
allowed to jump to a vacant site surrounded
by empty sites. It is necessary to have
a neighborhood (a set of one, two, or three sites,
depending on the model) of sites occupied. 

One advantage of using the conservative versions is
that they allow us to study the model 
without the danger of falling down into the absorbing state.
The conservation of particles permits us 
to perform numerical simulations that
avoid the accidental fall into the absorbing state.
Although they do not have absorbing states, they
are equivalent, in the thermodynamic limit, to the ordinary models.
The conservative and ordinary models are versions 
of the same model in distinct ensembles 
\cite{mario2003,tome2001,hilhorst,sabag};
the first models belonging to the constant-particle ensemble,
the second models belonging to the constant-rate ensemble.

Another advantage is related to the expected existence of 
first order transition. In the ordinary models, a very small 
change in the annihilation rate (the control parameter), 
near the transition, causes a great change in the density.
In the conservative models, because of the fact that the 
particle number is a conserved quantity (and therefore, works as
the control parameter), this problem does not occur. 
This advantage has been used by
Ziff and Brosilow \cite{brziff} in their 
study first order transition in the ZGB model. 

\section{Conserved reaction diffusion models}

In the construction of conserved models 
we have to be concerned only with the reaction process
since the diffusion process already conserves the particle number.
The necessary condition to set up an equivalent conserved version
of an arbitrary ordinary reaction-diffusion process in a lattice
is that the reaction process be a sum of a creation subprocess
and an annihilation subprocess \cite{mario2003}. 
This is always possible do realize
because these two subprocesses are mutually excludent.
If a site of the lattice is empty only creation is possible; 
if it is occupied, only annihilation is possible.
Therefore, the transition rate $w_i$
related the creation-annihilation of a particle at site 
$i$ can always be written in the form
\begin{equation}
w_i=k_C \omega_i^C + k_A \omega_i^A,
\label{0}
\end{equation}
where the first term is related to the creation of a particle at site
$i$ and the second to annihilation of a particle at site $i$.
The quantity $\omega_i^C$ vanishes 
if there is already a particle at site $i$ and 
$\omega_i^A$ vanishes if site $i$ is empty.
The quantities $k_C$ and $k_A$ are the actual parameters
of the ordinary model which we call amplitudes of the
creation and annihilation rates, respectively.

The conserved version is set up by replacing both the
creation and annihilation subprocesses by a particle jump process $i\to j$
with rate $w_{ij}=\omega_i^A \omega_j^C/L$ where $L$ is the
number of sites of the lattice. 
One can prove \cite{mario2003} that a two site
process defined by this transition rate
is equivalent in the thermodynamic limit to the ordinary process. 
To see how this come about let us look at the
total rate $\sum_i w_{ij}$ in which particles jump to site $j$. 
In the thermodynamic limit, the 
sum $\sum_i \omega_i^A/L$ approaches, by the law of large numbers,
the average $\langle \omega_i^A\rangle$
so that $\sum_i w_{ij} = \langle \omega_i^A\rangle \omega_j^C$.
By an analogous argument the total rate in which particles
leave the site $i$ is
$\sum_j w_{ij} = \langle \omega_j^C\rangle \omega_i^A$.
The averages $\langle \omega_i^A\rangle$ and 
$\langle \omega_i^C\rangle$ act then as the amplitudes of creation
and annihilation rates, respectively, what 
allows us to write down the following relation \cite{mario2003}
\begin{equation}
\frac{k_A}{k_C} = 
\frac{\langle \omega_j^C\rangle}{\langle \omega_i^A \rangle},
\label{tl}
\end{equation}
between the amplitude rates of the constant-rate
ensemble and averages determined in the constant-particle ensemble.

For the model we study here particles are spontaneous annihilated
do that $\omega_i^A$ is 1 if site $i$ is occupied and 0 if it is empty. 
Therefore $\sum_i \omega_i^A=n$ where $n$ is the number of particles. 
The quantity $\omega_j^C$ is 0 if site $j$ is occupied.
Since creation is catalytic, this quantity
depends also on the neighborhood of site $j$.
For the single-creation model it equals half the number of
nearest neighbor occupied sites. For the pair-creation
model it equals half the number of pairs of nearest neighbor 
occupied sites. For the triplet-creation model it equals half 
the number of triplets of nearest neighbor occupied sites.
It is convenient do define an active site as the site for which
$\omega_j^C$ is nonzero. The number of active sites $n_{\rm ac}$ 
is defined by 
\begin{equation}
n_{\rm ac}=\sum_j w_j^C.
\end{equation}
We also define a quantity $\alpha$ as being the right-hand side
of equation (\ref{tl}) so that, for the models studied here,
\begin{equation}
\alpha=\frac{ \langle  n_{\rm ac} \rangle }{n}.
\label{5}
\end{equation}
where the averages are taken in the constant-particle ensemble.
Usually one defines the ordinary reaction process so that the
rate amplitudes are $k_C=1$ and $k_A=k$. 
Therefore, according to relation (\ref{tl}), $\alpha$ 
coincides with the parameter $k$ of the ordinary model
as long as the average density of particles 
of the ordinary models
equals the density of particle $n/L$ in the conserved models.

The rules of the reaction-diffusion processes
we used are such that the diffusion occurs with probability
$D$ and the jump process with probability $1-D$. 
The quantity $D$ and the diffusion rate
$\tilde D$ are related by
\begin{equation}
D=\frac{\tilde{D}}{1+\tilde{D}}. 
\end{equation} 

\section{Exact and mean-field results}

The average number of active sites per site of the lattice
equals the probabilities $P(10)$, $P(110)$, and $P(1110)$ 
for the single-creation, pair-creation and triplet-creation models,
respectively. Since the number o particles per site is
the probability $P(1)$ it follows that $\alpha$ is given,
respectively, by 
$\alpha = P(10)/P(1)$,
$\alpha = P(110)/P(1)$, and
$\alpha = P(1110)/P(1)$, for the three models.
In the limit of infinite diffusion rate the particles will
be uncorrelated so that $P(10)=P(1)P(0)$, etc.
Taking into account that $P(1)=\rho$ and $P(0)=1-\rho$, 
we get the following exact results for the active state,
valid for $D=1$,
\begin{equation}
\alpha = 1-\rho,
\label{sc}
\end{equation}
for the single-creation model,
\begin{equation}
\alpha =\rho (1-\rho),
\label{pc}
\end{equation}
for the pair-creation model, and
\begin{equation}
\alpha = \rho^2 (1-\rho),
\label{tc}
\end{equation}
for the triplet-creation model.

These results give a continuous transition for the single-creation
model. For the pair-creation and triplet-creation models,
on the other hand, they give
a discontinuous transition since $\rho$ does not 
vanishes continuously as one increases $\alpha$.
The quantity $\alpha$ has a maximum at a nonzero value $\rho_0$
of the density which is $\rho_0=1/2$ for the pair-creation model and
$\rho_0=2/3$ for the triplet-creation. 
The corresponding
values of $\alpha$ are $\alpha_0=1/4$ and $\alpha_0=4/27$,
respectively. 
Since there is no free energy from which we could decide
at what point the jump in the density occurs
one is tempted to use the maximum value of $\alpha$
(spinodal point).
However, as we will see, our numerical results do not
support this point of view. According to the numerical results,
the discontinuity occurs at a smaller value of $\alpha$.

For the single creation the diffusion does not change the
nature of the transition. Even at infinite diffusion rate 
the transition is continuous as the exact result 
(\ref{sc}) shows. The critical line
on the diagram $D$ versus $\alpha$ can be obtained by a mean-field
approximation. By using a two-site mean-field approximation
we get a relation between $\rho$ and $\alpha$
which shows a continuous transition for all values of $D$ and
which recovers the exact result (\ref{sc}) when $D=1$.
The critical line obtained from this approximation is given by
\begin{equation}
D=\frac{2\alpha-1}{4\alpha - 2\alpha^2 -1},
\label{mf}
\end{equation}
showing that $\alpha\to1$ as $D\to1$ in accordance
with the exact result (\ref{sc}) and, as we will see, with
numerical simulations.

The exact results (\ref{pc}) and (\ref{tc}) 
for the pair-creation and triplet-creation models 
cannot be used to infer that the transition will remain
discontinuous for finite diffusion rate even if the rate is large.
An indication that the transition is continuous at low diffusion
and discontinuous for sufficiently large diffusion,
giving rise to a tricritical point,
comes from mean-field approximations which can be done
at several levels \cite{dick1986}. 
At the level of three sites the mean-field approximation
locates the tricritical point of the pair-creation model
at $D_t=0.032$ and $\alpha_t=0.1687$. For the triplet-creation
model it is necessary to use a higher order of approximation.
At the level of four sites the tricritical point occurs
at $D_t=0.017$ \cite{dictome}. Although both results are
in qualitative agreement with our numerical simulations
they are very low when compared with the figures coming from
the numerical simulations. 

\section{Numerical simulations}

We have simulated the conservative diffusive contact process
in a one-dimensional lattice. 
The actual simulation is performed as follows. 
At each time step a particle is selected at random,
say a particle at site $i$,
and one of its neighboring sites is chosen randomly,
say site $j$.
If this neighboring site is empty then we decide
which process to perform: the diffusion of particles, 
occurring with probability $D^\prime$, or the
creation-annihilation process, occurring with probability
$1-D^\prime$. If the diffusion process is chosen then the particle
at $i$ hops to the neighboring site $j$. 
If the creation-annihilation process is
chosen then any another particle of the system,
including the one at site $i$, is chosen
randomly and placed at site $j$.
In the case of the pair-creation or 
triplet-creation models, however, this only happens
if the chosen particle $i$ has at least
one or two nearest neighbor occupied sites, respectively. 
The relation between the probability $D^\prime$ we use
in the simulation and the actual probability of diffusion $D$
is $D^\prime=2D/(1+D)$ \cite{dick1990}. This is so because
we are choosing a particle from a list of occupied sites
and then choosing with equal probability one of the
neighboring sites to place the particle.

\subsection{Supercritical regime}

%----------------------- figure 1 -----------------------
\begin{figure}
\setlength{\unitlength}{1.0cm}
\includegraphics[scale=0.66]{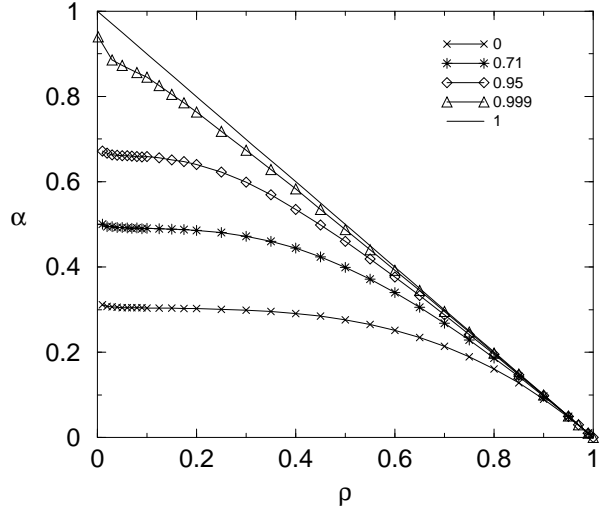}
\caption{\small{The effective number of active sites per 
particle $\alpha$ as function of  particle density $\rho$ for 
single-creation conservative contact process for some values 
of probability D.}}
\label{supsingle}
\end{figure}
%--------------------------------------------------------

%----------------------- figure 2 -----------------------
\begin{figure}
\setlength{\unitlength}{1.0cm}
\includegraphics[scale=0.66]{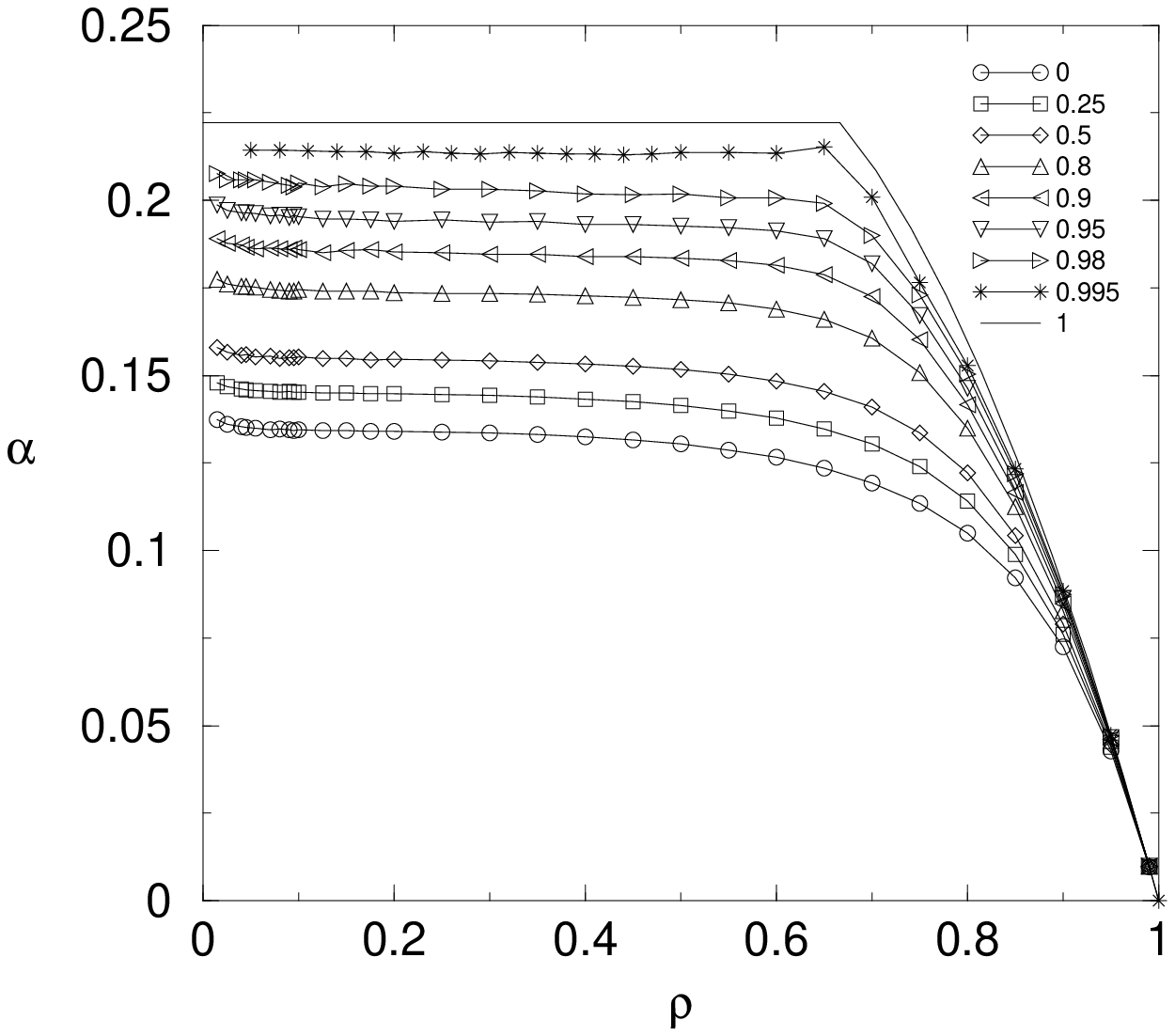}
\caption{\small{The effective number of active sites 
per particle  $\alpha$ as function of the particle density $\rho$ for
pair-creation  
conservative contact process for some values of probability D. 
The horizontal straight line at $\alpha =0.222$ was obtained 
by extrapolation.}}
\label{suppair}
\end{figure}
%--------------------------------------------------------

%----------------------- figure 3 -----------------------
\begin{figure}
\setlength{\unitlength}{1.0cm}
\includegraphics[scale=0.66]{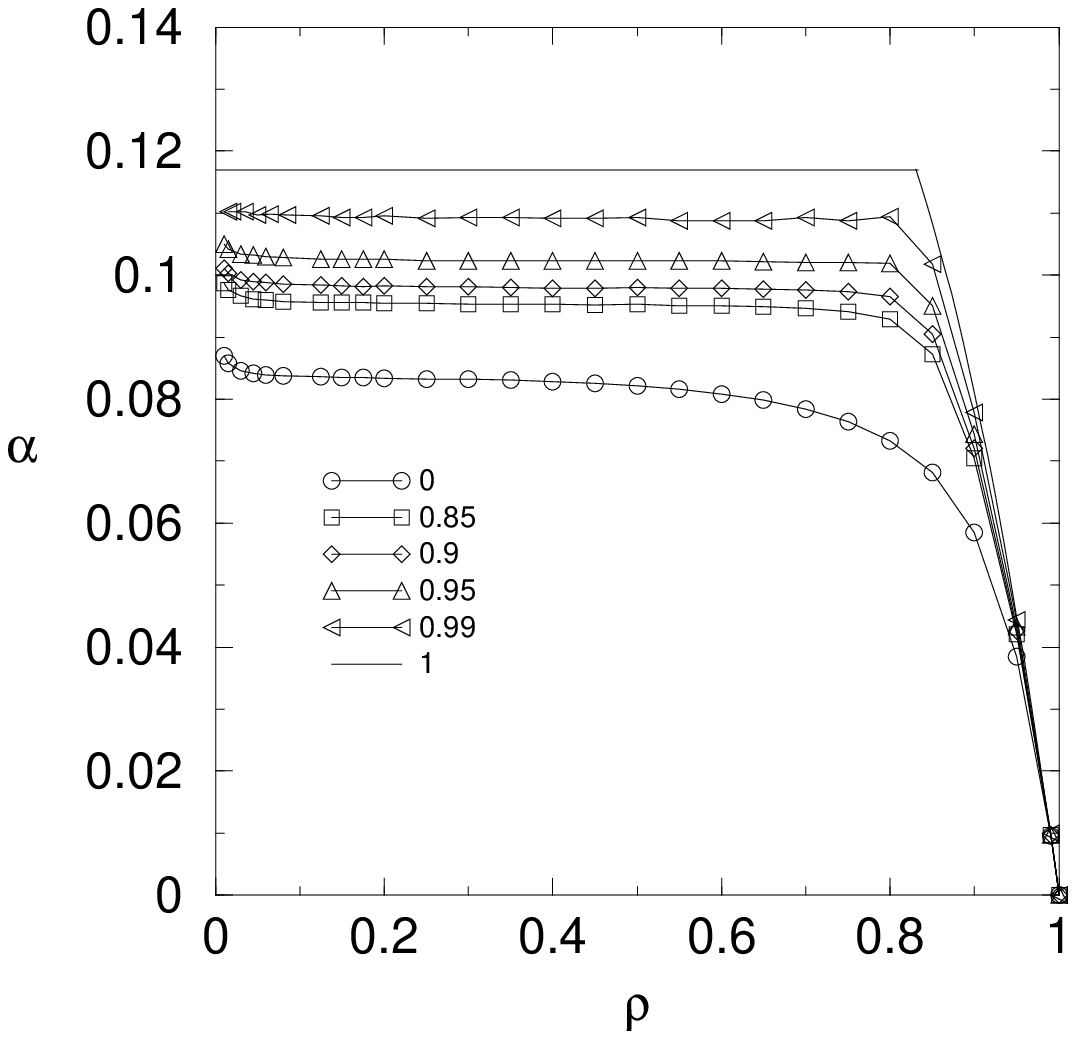}
\caption{\small{The effective number of active sites 
per particle $\alpha$ as function of the particle density $\rho$ for
triplet-creation  
conservative contact process for some values of probability D. 
The horizontal straight line at $\alpha =0.115$ was obtained 
by extrapolation.}}
\label{suptriplet}
\end{figure}
%--------------------------------------------------------

%----------------------- figure 4 -----------------------
\begin{figure}
\setlength{\unitlength}{1.0cm}
\includegraphics[scale=0.55]{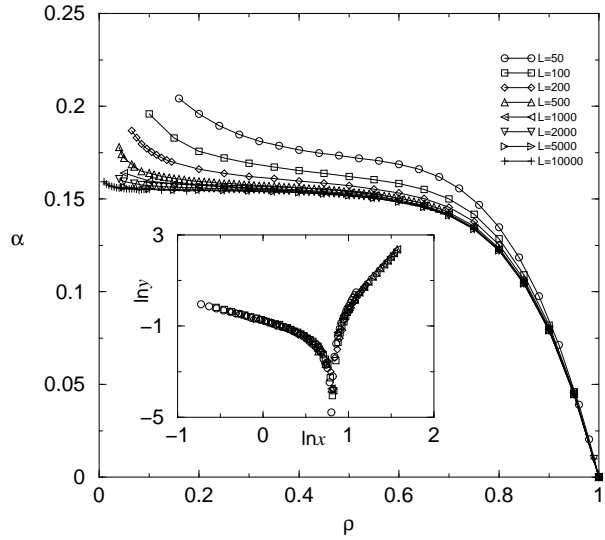}
\caption{\small{The effective number of active sites per particle  
$\alpha$ as function of the  particle density $\rho$ for
several values of $L$  in the supercritical regime for $D=0.5$ for the
pair-creation model. The inset show the scaling plot of
$y=L^{\beta/\nu}|\alpha_c-\alpha|$ versus $x=L^{1/\nu}\rho$, using the DP
critical exponents $\beta=0.277$ and $\nu=1.097$.
}}
\label{fig4}
\end{figure}
%--------------------------------------------------------

%----------------------- figure 5 -----------------------
\begin{figure}
\setlength{\unitlength}{1.0cm}
\includegraphics[scale=0.66]{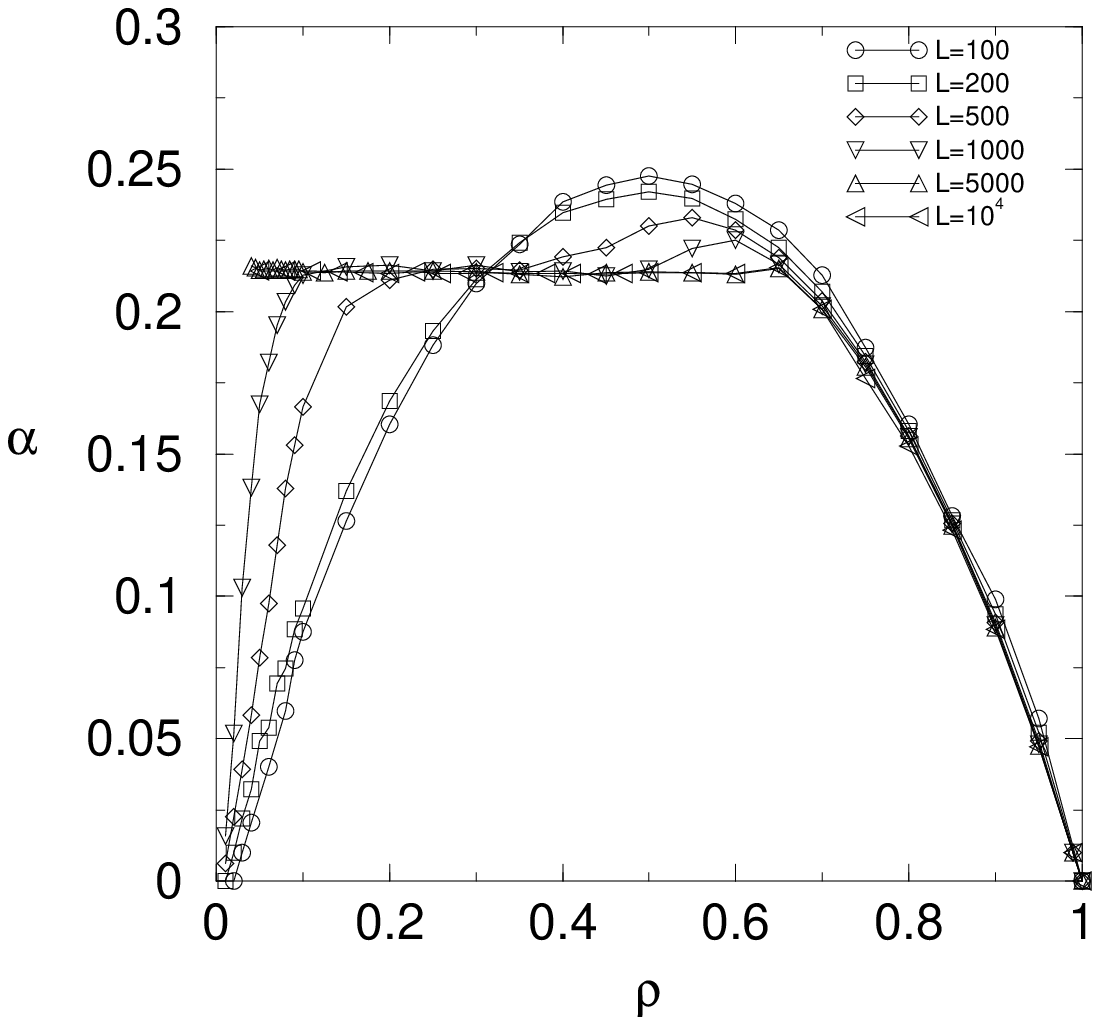}
\caption{ {The effective number of active sites per particle 
$\alpha$ as function of the  particle density $\rho$ for
several values of $L$  in the supercritical regime for $D=0.995$ for the
pair-creation model.}}
\label{fig5}
\end{figure}
%--------------------------------------------------------

The simulation was performed using lattices
with $L$ sites and periodic boundary conditions.  
The averages of the appropriates quantities were
obtained from a number of Monte Carlo steps
ranging from $10^6$ to $3\times10^7$, 
after discarding a sufficient number of steps
to reach the stationary state.
In Figs. \ref{supsingle}, \ref{suppair} and \ref{suptriplet}, 
we show the  particle density $\rho=n/L$
as a function of $\alpha$, calculated by using 
formula (\ref{5}), for several values of the hopping probability $D$. 
We have used $L=10^4$ and varied the number of particles $n$.
As expected, 
for high diffusion rate the curves approach the exact behaviors
given by equations (\ref{sc}), (\ref{pc}) and (\ref{tc}).

For the diffusive single-creation contact process, 
the transition is found to be continuous for all values of $D$.
Increasing the diffusion
probability $D$ the critical value of $\alpha$
increases towards the value $1$ when $D\to 1$
as expected. 
For the diffusive pair-creation and triplet-creation
models the phase transition is continuous for low diffusion 
becoming discontinuous for high enough diffusion.
The tricritical point occurs at $D_t=0.965 \pm 0.010$  for the pair-creation
contact process and   $D_t=0.945\pm 0.005$ for the triplet-creation
as we shall see shortly. 
Figs. \ref{suppair} and \ref{suptriplet} 
seems to show that this is indeed the case.

To compare the behaviors corresponding to the 
second and first order transitions, we 
simulated the pair-creation model at
$D=0.5$ and $D=0.995$ for various
values of the system size $L$ ranging from $50$ to $10^4$.
For the former case, $D=0.5$,
the plot of $\rho$ versus $\alpha$,
shown in Fig. \ref{fig4}, shows a continuous transition.
That the transition is continuous is confirmed by the 
the data collapse of the data shown in the inset of
Fig. \ref{fig4}. 
For $D=0.995$, the plot of $\rho$ versus $\alpha$,
shown in Fig. \ref{fig5}, displays a jump
when $L\to\infty$ increases.
Results similar to those of Figs. \ref{fig4} and \ref{fig5} are also
found for the triplet-creation model.

\subsection{Subcritical regime}

\begin{figure}
\setlength{\unitlength}{1.0cm}
\includegraphics[scale=0.66]{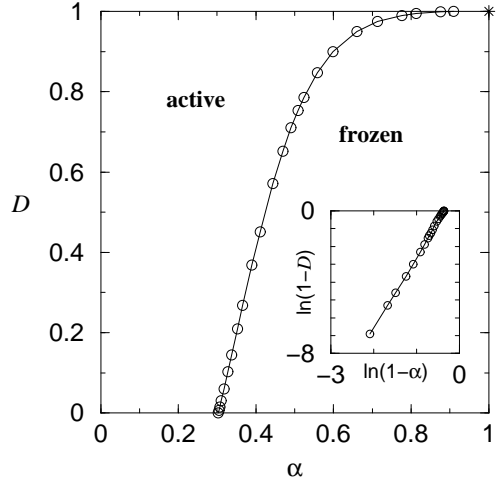}
\caption{\small{Phase diagram for the single-creation conservative 
contact process.  The star corresponds the value of $\alpha_{c}=1$ 
in the limit $D$ =1. The inset corresponds  the Log-log of 
equation (\ref{Log-Log}). 
The transition from active to frozen state is always second-order.}}
\label{pdsingle}
\end{figure}

\begin{figure}
\setlength{\unitlength}{1.0cm}
\includegraphics[scale=0.66]{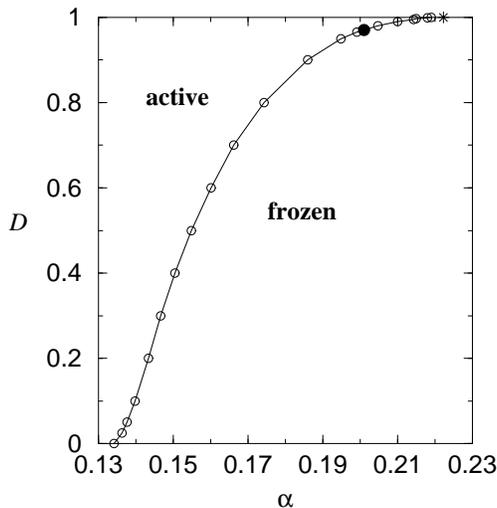}
\caption{\small{Phase diagram for the pair-creation conservative 
contact process. The star corresponds the value 
of $\alpha_{0}=0.222$ in the limit $D $ =1. The
tricritical point (full circle) is located  at 
$\alpha_{t}=0.199$ and $D_t=0.965$.}}
\label{pdpair}
\end{figure}

\begin{figure}
\setlength{\unitlength}{1.0cm}
\begin{picture}(-1.3,7.0)(4.15,1.)
\thicklines
\includegraphics[scale=0.66]{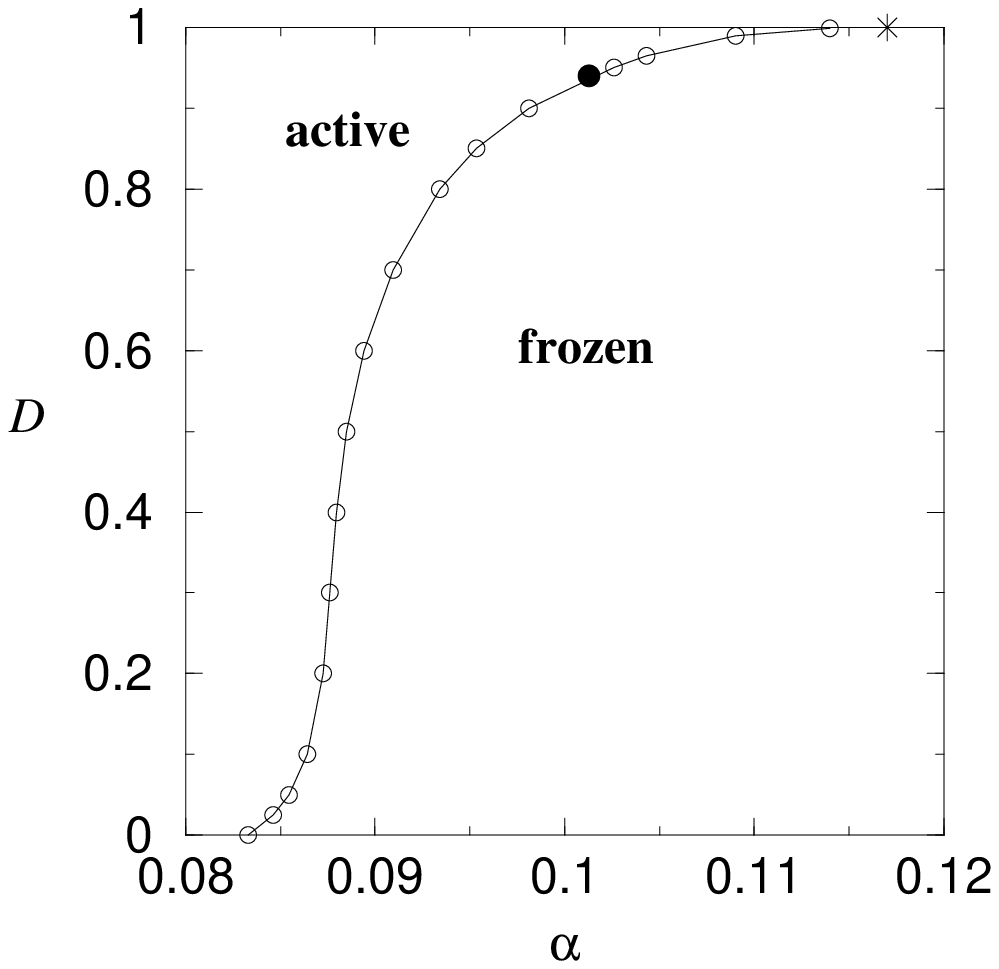}
\end{picture}
\caption{\small{Phase diagram for the triplet-creation conservative 
contact process. The star corresponds the 
value of $\alpha_{0}=0.115$ in the limit $D=1$ The
tricritical point (full circle) is located  at 
$\alpha_{t}=0.102$ and $D_t=0.945$. 
}}
\label{pdtriplet}
\end{figure}

To simulate the system in the subcritical
regime we consider an infinite lattice
with a finite number $n$ of particles.
In practice we use a finite lattice
and check whether a particle
reaches the border. If a particle
is about to reach the border we increase
the size of the lattice. For a fixed
value of $D$ we have simulated the system
for several values of $n$, computing
$\alpha$ by using (\ref{5}). 
For each value of $D$, the critical
value $\alpha_c$ was obtained in the
limit $n\to\infty$ by a linear
extrapolation in $1/n$. Using these
results we have built the phase diagram  
in the plane $D$ versus $\alpha$, as 
shown in Figs. \ref{pdsingle}, \ref{pdpair} and \ref{pdtriplet}.
The numerical values we have obtained for the transition line
agrees very well with the results obtained previously for
the ordinary models \cite{jen1993,dictome}.

%----------------------- figure 6 -----------------------
%--------------------------------------------------------

%----------------------- figure 7 -----------------------
%--------------------------------------------------------

%----------------------- figure 8 -----------------------
%--------------------------------------------------------

When $D\to 1$ the critical value of $\alpha$ 
approaches a limiting value $\alpha_0$.
For the single-creation model $\alpha_0 = 1$
as expected from the exact result (\ref{sc})
and also from the mean-field result (\ref{mf}).
Assuming that the behavior of $D$ around
$\alpha=1$ is given by
\begin{equation}
\label{Log-Log}
(1-D) \sim (1-\alpha)^{\phi},
\end{equation}
we have found from the plot shown in the inset of Fig. \ref{pdsingle} 
that $\phi=4.03(3)$. Notice that the mean-field behavior,
given by (\ref{mf}), predicts the value $\phi=1$.

\begin{figure}
\setlength{\unitlength}{1.0cm}
\begin{picture}(-1.3,7.0)(4.15,1.)
\thicklines
\includegraphics[scale=0.66]{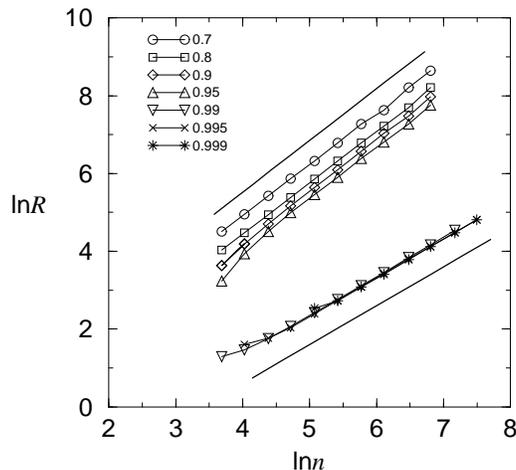}
\end{picture}
\caption{\small{Log-log plot of size of the system $R$, 
in the subcritical regime, as function
of the number of particles $n$ for several values of probability $D$ for the
pair-creation conservative contact process.  The upper straight line has slope
1.33 and the lower one has slope 1.}}
\label{lnpair}
\end{figure}

\begin{figure}
\setlength{\unitlength}{1.0cm}
\begin{picture}(-1.3,7.0)(4.15,1.)
\thicklines
\includegraphics[scale=0.66]{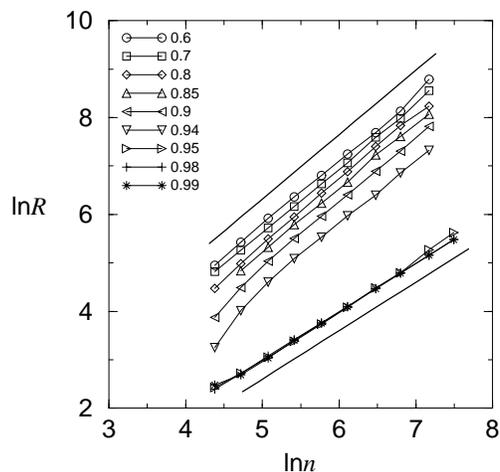}
\end{picture}
\caption{\small{ Log-log plot of size of the system $R$, 
in the subcritical regime, as function
of the number of particles $n$ for several values of probability $D$ for the
triplet-creation conservative contact process. The upper straight line has
slope 1.33 and the lower one has slope 1. }}
\label{lntriplet}
\end{figure}

An important feature of the models studied here
is the emergence of a fractal structure at 
the transition point, characterized by its fractal dimension.
We have calculated the fractal dimension
at the transition for each value of $D$. 
To this end we have simulated
a system with $n$ particles and determined
the average distance $R$ between the two 
particles located at the extremities of the system.
We assume the asymptotic behavior \cite{brograss}
\begin{equation}
\label{dimfrac}
n \sim R^{d_F},
\end{equation}
where $d_F$ is the fractal dimension,
so that the slope of a log-log plot of $N$ versus $R$
gives the fractal dimension as shown in 
Figs. \ref{lnpair} and \ref{lntriplet}, for the pair and
triplet-creation models, respectively.

%----------------------- figure 9 -----------------------
%--------------------------------------------------------

%----------------------- figure 10 ----------------------
%--------------------------------------------------------

For a continuous phase transition studied here we
expect the emergence of a fractal cluster
with a fractal dimension $d_F$ 
strictly less than one.
Indeed for the single-creation model and for the 
pair and triplet up to a certain value of $D$
we found a fractal dimension $d_F=0.75$ 
which is the expected value for 
a system in the DP universality class.
However, for the pair and triplet creation 
and for sufficient large values of $D$, the fractal
dimension becomes the Euclidean
dimension $d_F=d=1$ reflecting the
formation of a compact cluster whose
size $R$ increases linearly with $n$.
The changing of behavior occurs at 
$D_t=0.965\pm0.010$ and $\alpha_t=0.199\pm0.003$
for the pair-creation model and
$D_t=0.945\pm0.005$ and $\alpha_t=0.102\pm0.001$ for the triplet-creation
model. 
The tricritical point obtained by Dickman and Tom\'e \cite{dictome}
for the ordinary triplet-creation model by means of numerical
simulations is $D_t \simeq 0.85$ and $\alpha_t \simeq 0.096$.
These values correspond, actually, to a point over the critical line
in Fig. \ref{pdtriplet}. For $D=0.85$ our result is $\alpha=0.0954$.
As for the pair-creation model, the numerical results obtained by
Dickman and Tom\'e \cite{dictome} show that the transition is
continuous for $D<0.95$ which is consistent with our results.
However, they argue that the transition should remain continuous
for any finite diffusion.

We argue that the formation of a compact cluster
($d_F=1$) is a signature of a first order transition.
First of all, the compact
cluster has a nonzero density 
because $\rho=n/R$ does not vanish
in the limit $n\to\infty$
and should therefore be identified
with the active phase. Since the lattice is infinite
the active phase is in coexistence with
the frozen state (no particles). 
This behavior is very different from  that corresponding to a continuous
transition. In this case, the fractal
dimension is less than one which cannot be identified
with the active phase since
the density $\rho=n/R \sim n^{-(1-d_F)/d_F}\to 0$
when $n \to \infty$. 

When the cluster is a compact one, the ratio $n/R$ gives, in the limit
$n\to \infty$, the density $\rho_0$ of the active phase in coexistence
with the frozen phase. We have determined the values of $\rho_0$ 
for several values of $D$ above the tricritical point. An extrapolation
for $D=1$ gives $\rho_0=0.665(1) \approx 2/3$ for the pair-creation
model and $\rho_0=0.835(2) \approx 5/6$ for the triplet-creation model.
The values of $\alpha_0$ at the
first order transition can be obtained by substituting $\rho_0$
into the exact results (\ref{pc}) and (\ref{tc}). 
Using the numerical values, we get
$\alpha_0=0.222(3) \approx 2/9$ and 
$\alpha_0=0.115(1) \approx 25/216$ for the pair-creation model,
and the triplet-creation model, respectively.
As stated before, these values are distinct from the 
spinodal values coming from the exact solutions (\ref{pc}) and 
(\ref{tc}). We remark, on the other hand, that
the value of $\rho_0$ that we have obtained for the triplet-creation
model agrees with the value $\rho=0.84$
obtained by Dickman and Tom\'e \cite{dictome} for
the active coexistence phase at $D=0.95$.

\section{Conclusion}

The effect of diffusion in nonequilibrium systems has been
studied here for the case of three conservative contact processes.
For the single-creation contact process, 
the diffusion does not change the nature of
the phase transition, being continuous for any diffusion rate.
This is expected since the usual contact process has
already an intrinsic diffusion. Indeed, consider
the following sequence of transitions 
$010 \to 011 \to 001$ starting with an isolated particle.
The net result is a hopping of the isolated particle
to a neighboring site, or effectively a diffusion.
The sequence shown is a possible sequence of states
for the single-creation which is carried out by a 
creation followed by a annihilation.
This sequence, on the other hand, 
is not possible for the other two models.

For the pair-creation and 
and triplet-creation models the transition is continuous
for low diffusion and becomes discontinuous
for high enough diffusion. The present
approach in which the number of particles is conserved
is appropriate to study first order transition because it is possible to
distinguish this transition from a continuous one by measuring
the fractal dimension of the fractal cluster occurring at 
the critical point. If the fractal dimension is smaller
than the dimension of the lattice the transition is continuous.
When the cluster becomes compact, and the fractal dimension
equals the lattice dimension, the
transition becomes first order and, in addition,
the density of particles turns out to be the density
of the active phase in coexistence with the frozen phase.

\section*{ACKNOWLEDGMENT}
C. E. F. thanks the financial support from 
Funda\c c\~ao de Amparo \`a Pesquisa do
Estado de S\~ao Paulo (FAPESP) under Grant No. 03/01073-0.

%\begin{references}


\begin{thebibliography}{99}

\bibitem{harris} T. E. Harris, Ann. Probab. {\bf 2}, 969 (1974).

\bibitem{liggett} T. M. Ligget, {\it Interacting Particle Systems}
(Springer-Verlag, New York, 1995). 

\bibitem{marro}  J. Marro and R. Dickman, {\it Nonequilibrium Phase 
Transitions in Lattice Models} (Cambridge University Press, 
Cambridge, 1999).

\bibitem{dic89a} R. Dickman, Phys. Rev. B {\bf 40}, 7005 (1989).

\bibitem{dic89b}  R. Dickman, J. Stat. Phys. {\bf 55}, 997 (1989).

\bibitem{dic02} R. Dickman and M. A. F. de Menezes, 
Phys. Rev. E {\bf 66} 045101 (2002) .

\bibitem{jen1993}  I. Jensen and R. Dickman, 
J. Phys. A {\bf 26}, L151 (1993).

\bibitem{dictome} R. Dickman and T. Tom\'e, Phys. Rev. A {\bf 44},
4833 (1991).

\bibitem{brziff}  R. M. Ziff and B. J. Brosilow, Phys. Rev. A {\bf 46}, 4630
(1992).

\bibitem{brograss} H.-M. Br\"oker and P. Grassberger, Physica A {\bf 453}
 (1999).

\bibitem{tome2001} T. Tom\'e and M. J. de Oliveira, 
Phys. Rev. Lett. {\bf 86}, 5643 (2001).

\bibitem{hilhorst} H. J. Hilhorst and F. van Wijland,
Phys. Rev. E {\bf 65}, 035103 (2002). 

\bibitem{sabag} M. M. S. Sabag and M. J. de Oliveira, Phys. Rev. E
{\bf 66}, 036115 (2002).

\bibitem{mario2003} M. J. de Oliveira, 
Phys. Rev. E, {\bf 67 }, 027104 (2003).

\bibitem{dick1986} R. Dickman, Phys. Rev. A {\bf 34}, 4246 (1986).

\bibitem{dick1990} R. Dickman, Phys. Rev. A {\bf 42}, 6985 (1990).

%\end{references}
\end{thebibliography}
\end{document}